\documentclass[]{spie}  

\newcommand{\apix}{AstroPix}
\newcommand\apixone{AstroPix\_v1}
\newcommand\apixtwo{AstroPix\_v2}
\newcommand\apixthree{AstroPix\_v3}
\newcommand\amx{AMEGO-X}
\newcommand\um{$\mu$m}
\newcommand\gam{$\gamma$-ray}
 
\usepackage{amsmath,amsfonts,amssymb}
\usepackage{graphicx}
\usepackage[colorlinks=true, allcolors=blue]{hyperref}
\usepackage{caption}

\title{AstroPix: Novel monolithic active pixel silicon sensors for future gamma-ray telescopes}

\author[a,b]{Amanda L. Steinhebel}
\author[c,b,d]{Henrike Fleischhack}
\author[e]{Nicolas Striebig}
\author[f]{Manoj Jadhav}
\author[g]{Yusuke Suda}
\author[f]{Ricardo Luz}
\author[b]{Carolyn Kierans}
\author[b]{Regina Caputo}
\author[h]{Hiroyasu Tajima}
\author[e]{Richard Leys}
\author[e]{Ivan Peric}
\author[f]{Jessica Metcalfe}
\author[b]{Jeremy S. Perkins}

\affil[a]{NASA Postdoctoral Program Fellow}
\affil[b]{NASA Goddard Space Flight Center, Greenbelt, MD, USA}
\affil[c]{Catholic University of America, Washington, DC, USA}
\affil[d]{Center for Research and Exploration in Space Science and Technology, NASA/GSFC, Greenbelt, MD, USA}
\affil[e]{Karlsruhe Institute of Technology, Karlsruhe, Germany}
\affil[f]{Argonne National Laboratory, Lemont, IL, USA}
\affil[g]{Hiroshima University, Higashi-Hiroshima City, Hiroshima, Japan}
\affil[h]{Institute for Space-Environment Research, Nagoya University, Nagoya, Aichi, Japan}

\authorinfo{Further author information: (Send correspondence to A. Steinhebel)\\ A. Steinhebel: E-mail: \href{mailto:amanda.l.steinhebel@nasa.gov}{amanda.l.steinhebel@nasa.gov}}

\graphicspath{{./}{figures/}}
 
\begin{document} 
\maketitle

\begin{abstract}
Space-based \gam{} telescopes such as the $Fermi$ Large Area Telescope have used single sided silicon strip detectors to track secondary charged particles produced by primary \gam s with high resolution. At the lower energies targeted by keV-MeV telescopes, two dimensional position information within a single detector is required for event reconstruction - especially in the Compton regime. This work describes the development of monolithic CMOS active pixel silicon sensors – \apix{} – as a novel technology for use in future \gam{} telescopes. Based upon sensors (ATLASPix) designed for use in the ATLAS detector at the Large Hadron Collider, \apix{} has the potential to maintain high performance while reducing noise with low power consumption. This is achieved with the dual detection and readout capabilities in each CMOS pixel. The status of \apix{} development and testing, as well as outlook for future testing and application, will be presented.
\end{abstract}

\keywords{AstroPix, silicon pixel detectors, Compton telescopes, MAPS, high energy astrophysics }

\section{INTRODUCTION}
\label{sec:intro}  

The modern era of multi-messenger astronomy requires that many cosmic messengers be monitored, including the full electromagnetic spectrum. The `MeV Gap' in instrument sensitivity to \gam s in the keV-MeV range must be addressed to allow input to multi-messenger investigations in this regime. In order to ensure high precision and sensitivity, a next-generation Pair/Compton telescope targeting the MeV \gam{} sky requires detectors with high energy and position resolution in three-dimensions, as well as a low energy threshold. The technology used to track the paths of secondary charged particles resultant of incident \gam s in currently flying \gam{} telescopes such as single- and double-sided silicon strips cannot easily meet these challenges - the long strips introduce noise which impacts the measurement (see Ref~\citenum{Brewer:2021mbe}). Instead, Complementary Metal-Oxide-Semiconductor (CMOS) monolithic active pixel silicon (MAPS) sensors can be used. 

\begin{figure}
	\centering
	\includegraphics[width=0.59\linewidth]{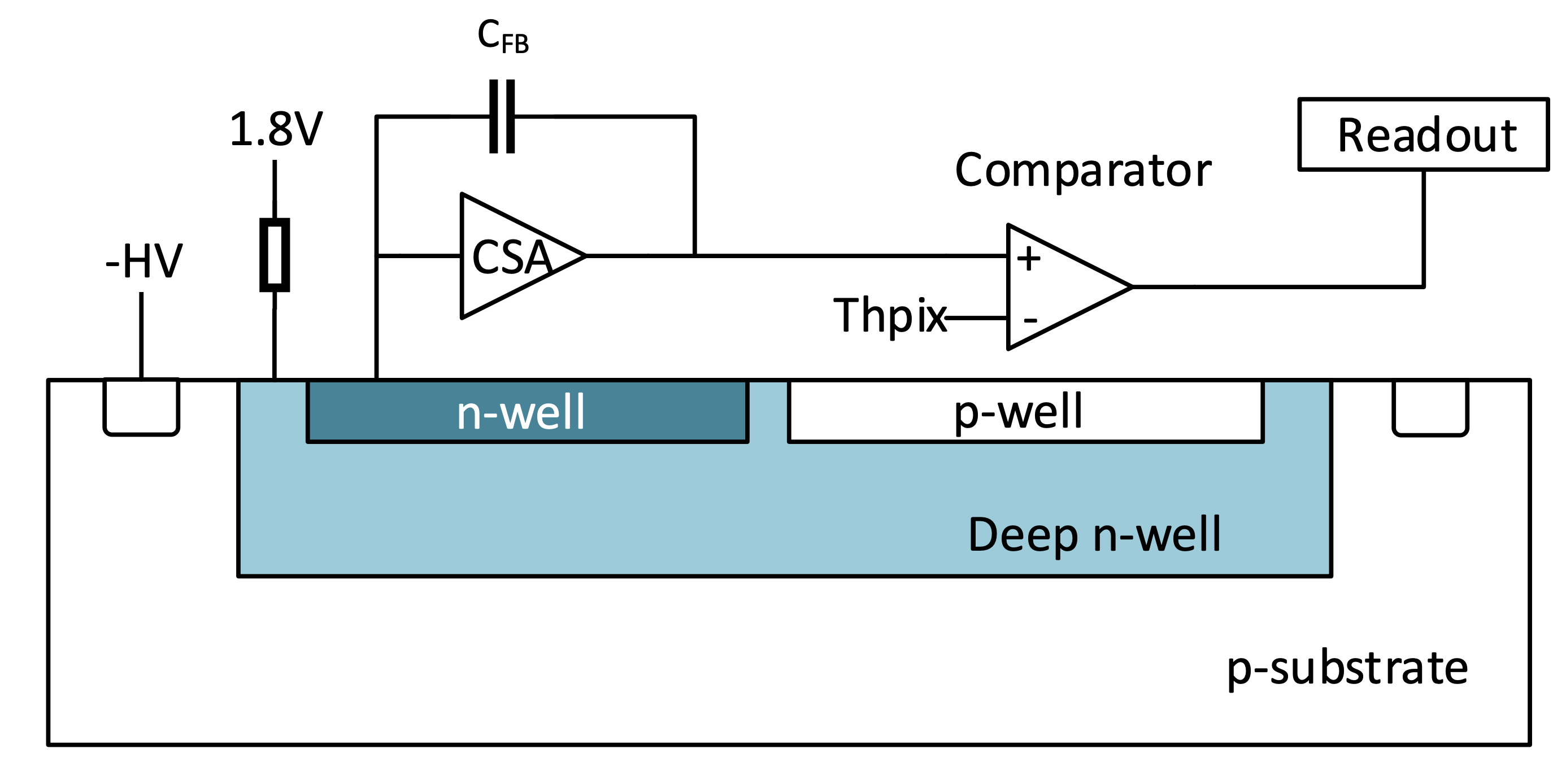}
	\includegraphics[width=0.39\linewidth]{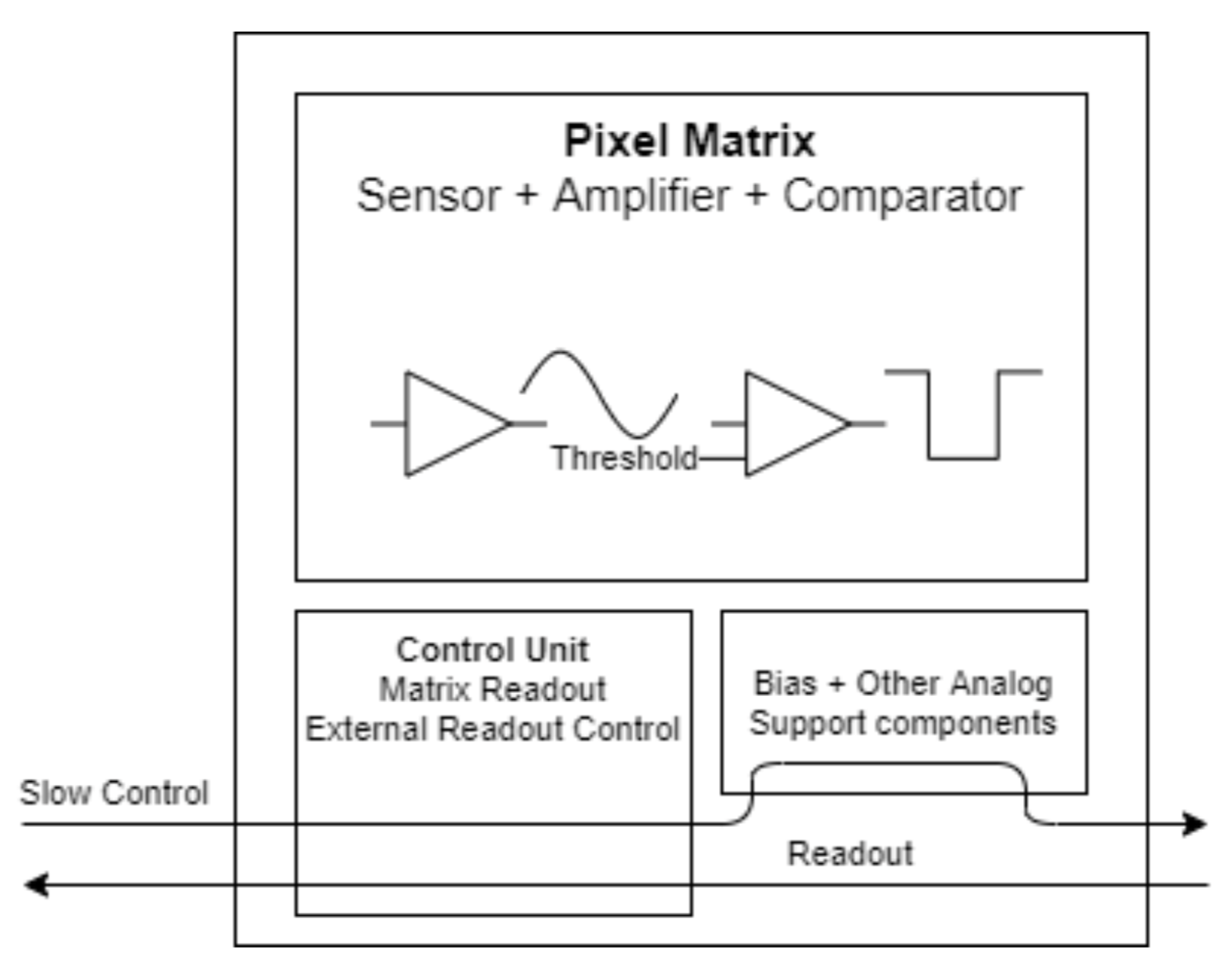}
	\caption{Illustration of an HVMAPS CMOS pixel with HV bias voltage (left) containing charge collection and amplification/comparator architecture embedded in the pixel matrix (figure from \citenum{striebig}). This matrix reads out to a control unit on the digital periphery (right) for digitization and chip readout.} 
	\label{fig:cmos}
\end{figure}

Individual pixels on these sensors can a range in size from 10~\um$^2$ - 1~mm$^2$, allowing for great control over position resolution (though this must be optimized with respect to power draw and data rate). The use of in-pixel CMOS circuitry (see Ref.~\citenum{Peric:2018lya}) eliminates the necessity for an external ASIC to record and digitize data as the charge collection and digitization is all done on-chip. A CMOS chip (see Fig.~\ref{fig:cmos}) incorporates charge collection, signal amplification, and a comparator for signal acceptance in a shared substrate embedded directly into the pixel matrix in shallow wells. The wells are isolated from the bulk substrate by a deep n-well. A charge-depleting bias voltage is applied individually to every pixel allowing for faster and more complete charge collection as compared to sensors that rely on diffusion. This makes AstroPix an HVMAPS sensor - MAPS that are depleted with high-voltage (HV) ($\mathcal{O}(100$V)) at a pixel level.

The comparator output for every pixel is sent to the synthesized digital logic which is located on the periphery of the chip. Here, the signal is digitized and can be read out for analysis. In this way, there is no need to directly bump-bond to each pixel to achieve high position resolution (as done with hybrid technology) as this strategy would introduce passive material into the active area of the detector. Large detectors designed with silicon strips require daisy-chaining the strips together, but the lack of this daisy-chaining with a HVMAPS sensor reduces capacitance and therefore noise in the signal. The AstroPix data collection strategy aims to reduce per-pixel power and overall data bandwidth. To this end, the comparator outputs for each row and column are OR'ed together and these two channels are sent to the digital periphery. In this way, the output signal is similar to the output of a silicon strip-based detector and row/column pairing algorithms will be employed to determine which signals correspond to which interaction if there are multiple triggers in a chip. Currently, AstroPix utilizes SPI readout from a DAQ which interfaces with an FPGA.

CMOS technology is familiar to industry and commercial endeavors, but novel in this application for space-based telescopes. \apix{} is a CMOS HVMAPS under design for this explicit purpose. The design of \apix{} is based upon work previously conducted by the ATLAS Collaboration at the Large Hadron Collider (LHC) for the development of an upgrade to their inner tracker subdetector system (see Ref.~\citenum{schoning2020mupix}). This custom MAPS detector, ATLASPix, was designed for the LHC environment - optimized for minimum ionizing particles with nanosecond timing capability. 

The \apix{} design philosophy utilizes ATLASPix as a starting point, and involves a sequence of \apix{} versions in order to advance the design in steps from ATLASPix to a flight-ready \apix. These versions are denoted with an underscored \verb|_v|. This strategy ensures functional chips at each stage of testing. With continued optimization and characterization studies such as those outlined in this work, \apix{} can serve as a baseline detector for future telescope concepts - especially those that require high-resolution and low-noise/low-power electronics or those targeting the keV-MeV \gam{} regime. For more background information about detection strategies of keV-MeV \gam s and its impact on telescope design, see Ref.~\citenum{Brewer:2021mbe}.

Section~\ref{sec:v1} outlines the first changes made to ATLASPix to produce a $bona-fide$ \apix{} chip, \apixone. This section overviews the characterization effort of this sensor, including analysis of analog data, background estimation, and the creation of a calibration curve. Following the design strategy of incremental changes, \apixone{} was superseded by \apixtwo {} - the current design under testing. Section~\ref{sec:v2} mirrors the structure of Section~\ref{sec:v1} with the updated sensor. \apixtwo{} was exposed to multiple particle beams and underwent preliminary radiation testing which is discussed in Section~\ref{sec:rad}. Section~\ref{sec:future} outlines plans for the next version of \apix, \apixthree, and environmental testing plans. It also contains a general outlook and summary.

\section{ASTROPIX$\_$V1}
\label{sec:v1}

The feasibility of using ATLASPix as a technological starting point for AstroPix was directly tested at NASA Goddard Space Flight Center (GSFC) in Ref.~\citenum{Brewer:2021mbe} where the measured energy resolution was found to be 7.69$\pm0.13\%$ at 5.89 keV and $7.27\pm1.18\%$ at 30.1 keV. Analog recorded data (response pulses as measured from one pixel) was used for this measurement. Digital data (bitstreams with encoded hit information from the full array) was also considered, but individual pixel responses were untuned resulting in large spreads in the Gaussian-fit response distribution (up to $\sigma/\mu=60$\%) when considering the summed response of the array as a whole.
 
Testing and characterization of the first \apix{} version, \apixone, has been completed and is presented in this section. Like ATLASPix, both analog data and digital data will be collected from \apix. Although digital data is intended to be utilized in flight, preliminary analog data is the focus of \apixone{} studies as digital data readout was not available in \apixone{} due to a flaw in the chip design. Missing shielding allowed parasitic capacitance to induce feedback and cause oscillations in the comparators. These oscillations were reproducible in simulation and are considered understood. All subsequent versions of the chip correct this defect.

Unless otherwise specified, the energy resolution is defined as the ratio of full width at half maximum (FWHM) of the reconstructed energy distribution over the mean of the reconstructed energy, expressed as a percent. The reconstructed energy spectrum for a mono-energetic input is assumed to follow a Gaussian shape, with mean $\mu$ and width $\sigma$. Thus, $E_{res} = 2.355*|\sigma|/\mu~*100\%$.

\subsection{Design Optimization for \apixone}
\label{sec:design_v1}

The sensitivity of future \gam{} telescopes in the keV-GeV range is driven by the effective area, angular resolution, and energy resolution of the instrument. By using a thick wafer of at least 500~\um, \apix{} must measure energy resolution at FWHM of 5~keV from 25-122 keV and 5.6 keV at 622 keV with a pixel dynamic range of 25-700 keV. \apix{} pixel size is a compromise between position resolution and power consumption. Smaller pixels provide higher position resolution 
however an increased number of pixels causes increased data rates and higher overall power consumption. 

ATLASPix provided a promising starting point, but itself does not meet telescope design criteria (see Table~\ref{tab:designCriteria}). From here, \apixone{} was designed with 18 rows and columns of $175\times175$~\um{} pixels on an un-thinned 725~\um{} thick wafer. The wafer thickness enables the larger dynamic range desired by \apix, and larger pixels do not sacrifice energy resolution. Loosening the timing requirement from ATLASPix allows for a simpler readout system. In turn, noise-reducing components could be added on-chip. For example, the addition of a low pass filter after the charge amplifier allowed for improved energy resolution. 

\begin{table}
	\caption{Parameters of ATLASPix and \apix.}
	\label{tab:designCriteria}  
	\begin{center}
	\begin{tabular}{|l|cc|}\hline  
	~ & ATLASPix & AstroPix \\
	~ &(Measured) & (Required) \\\hline
	Sampling rate [ns]& 16& $\sim$ 600\\ 
	$E_{res.}$ (FWHM)& 7\% at 30.1 keV& 9.7\% at 122 keV\\ 
	Pixel size [\um]& 150$\times$50& $1000\times1000$\\
	Thickness [\um]& 100& 500\\ 
	Dynamic range [keV]&5-32&25-700\\
	Power consumption [mW/cm$^2$]&150&1.5\\\hline
	\end{tabular}
	\end{center}
\end{table} 

\subsection{\apixone{} Testing Setup}
\label{sec:setup_v1}

\begin{figure}
	\centering
	\includegraphics[width=0.3\linewidth]{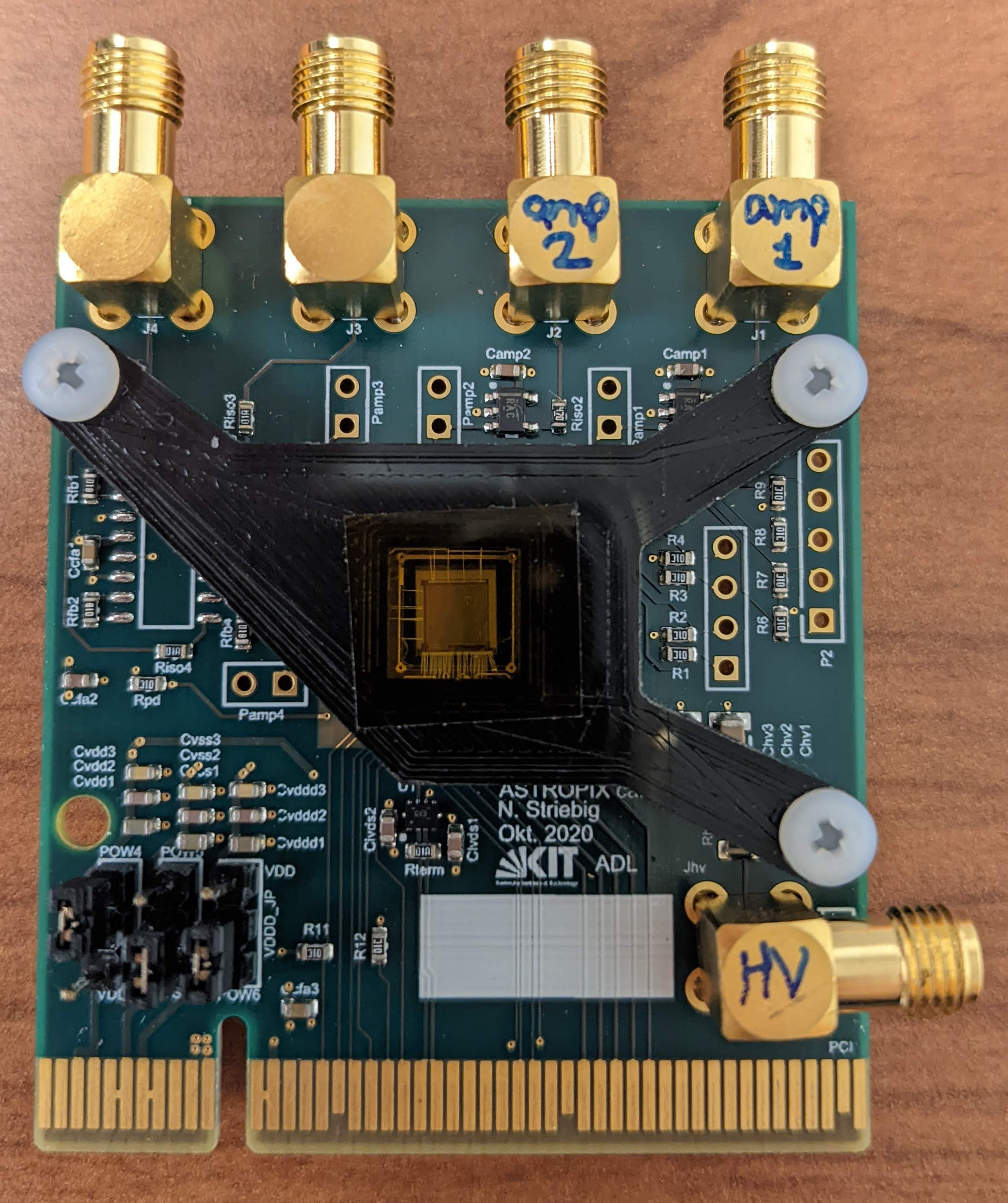}
	\caption{\apixone{} mounted on a carrier board. A 3D-printed shield covers the chip with a layer of kapton to protect the wireboards.}
	\label{fig:v1board}
\end{figure}

\apixone{} is controlled and configured by a Nexys Video FPGA and readout system adaptor (GECCO) board. 
Three extension boards are required, which enable 1) configuration, 2) voltage setting, and 3) the injection of a square analog transient signal for chip testing.
%
\apixone{} is mounted to a carrier board (see Fig.~\ref{fig:v1board}) using conductive pressure sensitive tape and then wire-bonded. This carrier board connects directly to the GECCO board through an PCIe slot. 
High-voltage reverse bias is supplied for charge depletion on a pixel-by-pixel basis. 

Analog signals are displayed on a mixed domain oscilloscope. SMA ports on the injection card and output pixels are connected directly to the oscilloscope. Two \apixone{} chips mounted to carrier boards were tested at GSFC and are considered in this work - Chip003 and Chip004. Each chip has 18 pixels connected to pads and capable of analog readout. Two pixels were chosen for the studies presented in this work - (1,18) or `amp1' and (18,1) or `amp2' indicates pixel (18,1).  This provides two pixels in the corners of the array to avoid crosstalk. Known chip defects prohibit the use of the digital readout system, so all studies performed in the work will focus only on the analog signal detected in these two spatially separated pixels.

Response signals are read with a supplied bias voltage of -60V. This value is not optimized for full depletion, and as such it is not anticipated that \apixone{} meets the full \apix{} design requirement of 500~\um{} depletion. With the $250\pm50~\Omega*$cm resistivity wafers used for \apixone, ithe depletion depth was estimated to 3-100~\um.

\subsection{\apixone{} Characterization Studies}

\subsubsection{Charge Injection Studies}
\label{sec:inj_v1}

\apix{} is designed along with the capability to receive an injected charge signal. This artificial signal is used as an initial test to confirm that correct configuration is successfully being passed to the \apix{} chip, and that analog and digital responses can be triggered upon and read out. The injected signal itself, a square wave with programmable frequency and duration can be observed with a digital oscilloscope. This injected pulse is used to trigger data collection upon the falling edge. 

\begin{figure}
	\centering
	\includegraphics[width=0.5\linewidth]{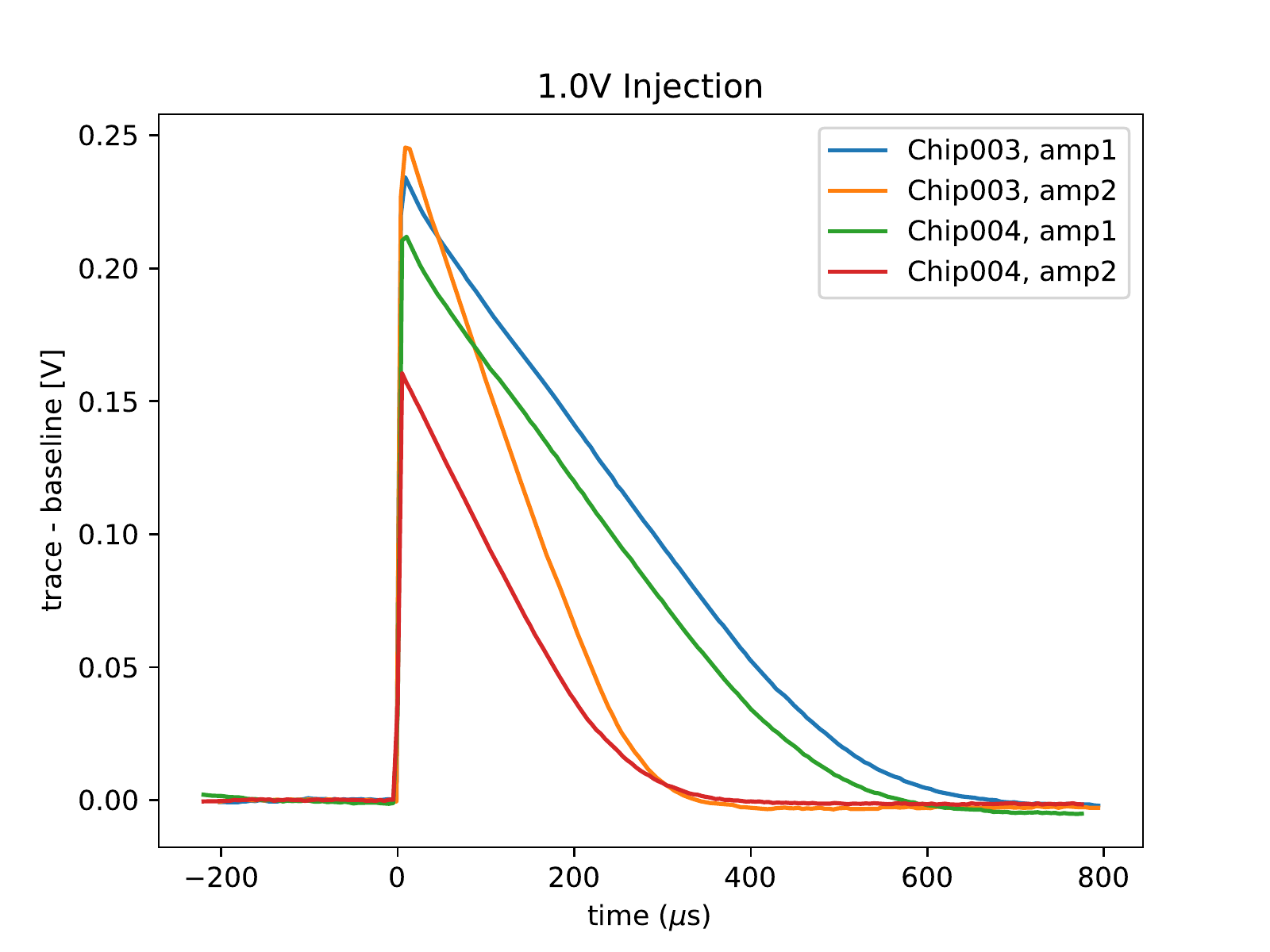}
	\caption{The average of 100 traces from both \apixone{} chips in response to an injected pulse of 1.0~V.}
	\label{fig:compareInjChips}
\end{figure}

Using a bias voltage of -60~V, both pixels respond but with different pulse shapes. For chip003, amp1 reliably has lower pulse height and longer pulse duration than amp2. This in part is due to the shaping seen in amp1 with a more linear pulse decay rather than the more exponential nature seen in amp2. This behavior is opposite for chip004 (see Fig.~\ref{fig:compareInjChips}) where amp1 showed much higher pulse heights than amp2. The linear or exponential nature of the pulse decay are consistent between amp1 and amp2 of the different chips. This leads to a pulse duration for amp1 of both chips of around 600~$\mu$s compared to 300~$\mu$s for amp2. As this analog signal is readout from the chip prior to digitization, the differences in pulse shape are related to the variability of amplification and shaping electronics present in each pixel.

From this study, we can conclude that the probed pixel both respond to the injected charge albeit with different pulse shapes. This builds confidence in the circuitry and response of the chip, as well as data readout, collection, and analysis methods. It also illustrates the degree of variation that can occur in pixel responses. In future \apix{} versions, pixels will be indiviually tuned to limit the variability in pulse shape.

\subsubsection{Threshold Measurements}
\label{sec:noise_v1}


\apix{} requirements define a 25 keV low threshold. This enables the measurement of low-energy astrophysical photons through photoabsorption into a single pixel. In order to measure the threshold of \apixone{} and compare the performance to requirements, a low threshold for data collection must be set such that the sensor is protected from randomly triggering upon noise at an overwhelming rate. 

Low-energy depositions from ambient sources (such as fluorescent lighting) or electronics noise are disregarded as background from subsequent analyses with a hard cut on acceptable pulse heights. This defines a low threshold with which measurements are recorded. Compton-scattered deposits within this regime cannot be uniquely identified and are lost. Dedicated runs with no radioactive source and a threshold low enough to trigger upon background hits are used to define the threshold level as that which eliminates all activity from the background runs. This threshold is set on the analog data collecting oscilloscope with millivolt precision. The shape of the resulting background distribution of hits is Gaussian. The amplitude of the response of the analog probed pixels in different chips can differ due to fabrication differences (see Fig.~\ref{fig:compareInjChips}). Minimum thresholds of analog signal pulse height for different pixels are shown in Table~\ref{tab:noiseThreshold}. These threshold values can also be interpreted as energy values after calibration is defined in Section~\ref{sec:calib_v1}.

Additional background events from Compton-scattered particles from radioactive sources cannot be removed with a hard cut. These are included in subsequent analysis. 

\begin{table}[b!]
	\caption{Threshold peak height required for a trace to be considered `signal' (and not background or noise).}
	\label{tab:noiseThreshold} 
	\begin{center}
	\begin{tabular}{|c|c|c|c||c|c|c|}\hline
		Version&Chip&Pixel&Threshold [mV]&Chip&Pixel&Threshold [mV]\\\hline
		v1&003&amp1&20&004&amp1&10\\
		v1&003&amp2&30&004&amp2&20\\
		v2&1&amp1&60&-&-&-\\
		\hline
	\end{tabular}
	\end{center}
\end{table}

\subsubsection{Energy Calibration}
\label{sec:calib_v1}

Charge injection studies of Sec.~\ref{sec:inj_v1} validate that \apixone{} responds to signals and fostered the development of software tools, including data collection and analysis. From here, a correlation between \apixone{} response (analog pulse height) and incident particle energy is measured in order to calibrate response spectra into units of energy. Radioactive isotopes with known emission lines were used. The sources utilized in these calibration studies (see Table~\ref{tab:sources}) have emission lines spanning 14 keV - 122 keV. These sources probe the low end of \apix 's sensitivity regime. Statistics in these studies can be limited due to the radioactivity of the source and the analog readout scheme. Analog data from only one single pixel can be read out during a collection run, leading to a very small sensitive area on the chip. Additionally, the cross section decays with increased photon energy, leading to lower statistics for higher energy incident particles.
\begin{table}[b!]
	\caption{Properties of radioactive isotopes used for calibration curve creation. The Compton edge of the Cobalt-57 122 keV photopeak is also considered.}
	\label{tab:sources} 
	\begin{center}
	\begin{tabular}{|l|c|c|}\hline
		Isotope&Radioactivity [mCi]&Emission Line(s) [keV]\\\hline
		Cobalt-57&0.9973&14.41, 122.06\\
		&&39.46\\
		Cadmium-109&1.031&22.16, 88.03\\
		Barium-133&1.05$\times 10^{-4}$&30.97\\
		Americium-214&0.1065&59.54\\
		\hline
	\end{tabular}
	\end{center}
\end{table} 

\begin{figure}
	\centering
	\includegraphics[width=0.6\linewidth]{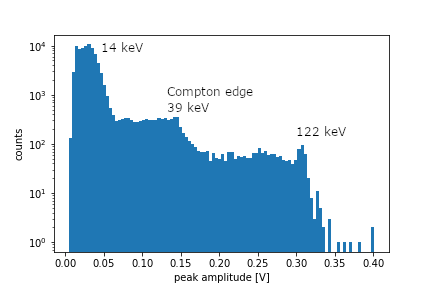}
	\caption{Spectrum of the raw data collected over 16 hours with \apixone{} chip003 amp1 of Cobalt-57. Photopeaks associated with 14.41 keV and 122.06 keV emission can be seen, as well as a Compton edge at 39.46 keV. Between these features, \apixone{} experiences high background rates from Compton-scattered interactions.}
	\label{fig:rawSpectrumExample}
\end{figure}

Analog data was triggered for readout by the \apixone{} response itself. The height of each response peak is utilized as a proxy for energy deposition. The collection of these signals create spectra of raw data (see Fig.~\ref{fig:rawSpectrumExample}) that contain peaks associated with anticipated photopeaks but also high rates of signal recorded between expected peaks. For the remainder of this work, the 30.97 keV photopeak of Barium-133 will be used a particular case study to illustrate raw and calibrated spectra, performance of calibration software, and \apix{} response. Like with Cobalt-57, below the Barium-133 photopeak backscattered signal can also be identified. 

\begin{figure}
	\centering
	\includegraphics[width=0.45\linewidth]{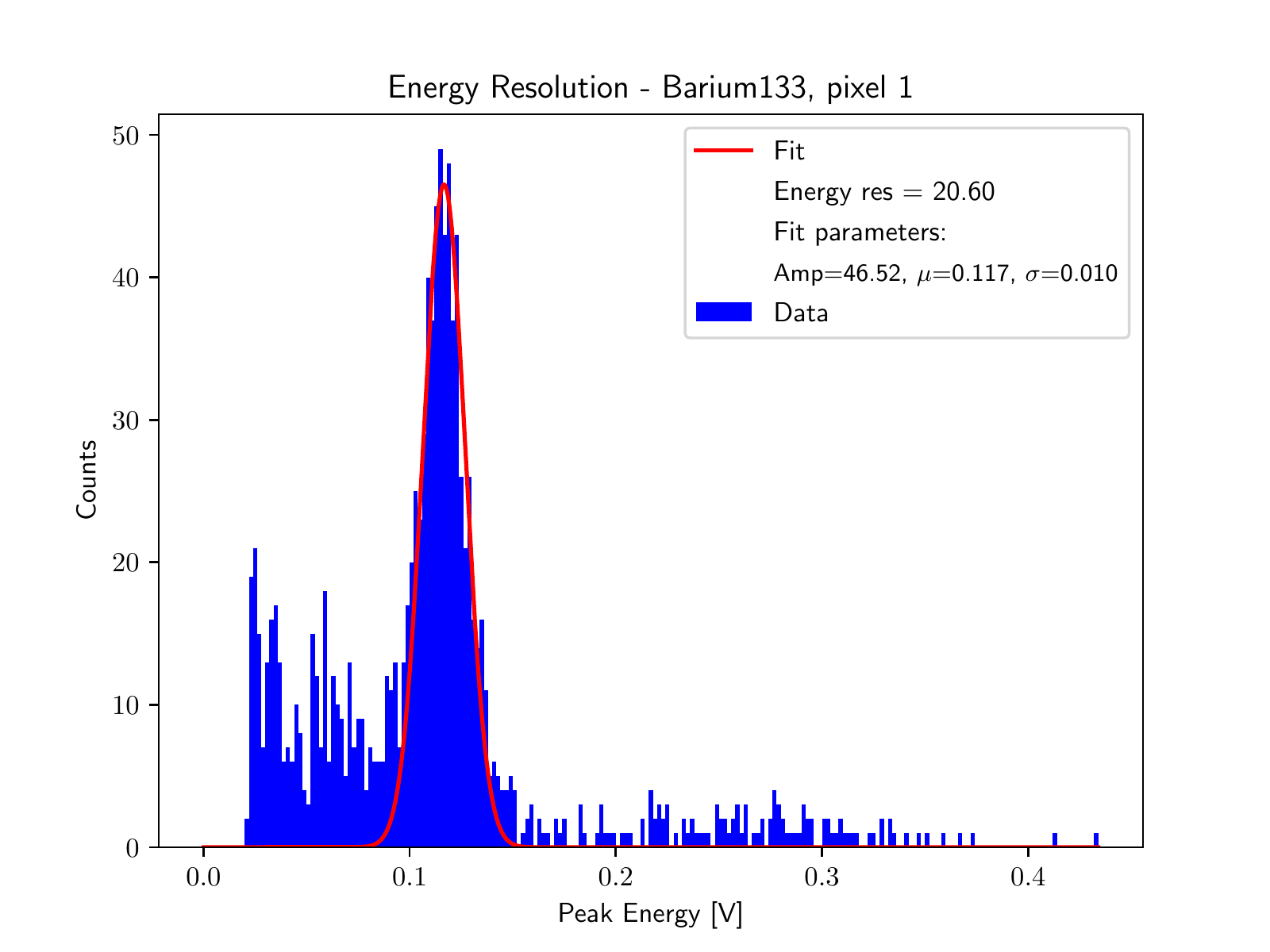}
	\includegraphics[width=0.45\linewidth]{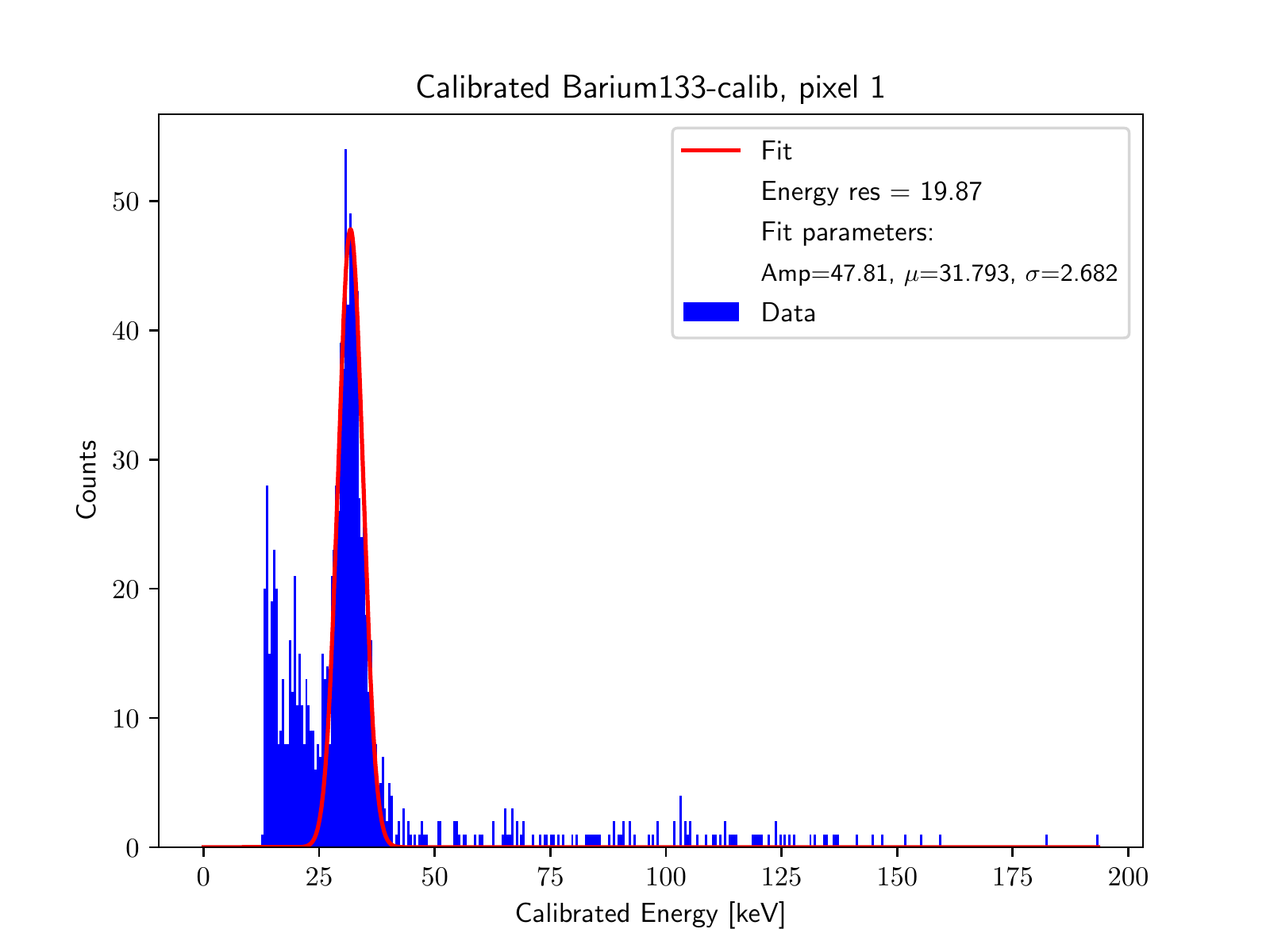}
	\caption{(Left) Gaussian fit to raw data collected with \apixone{} chip003 amp1 of the Barium-133 30.97 keV photopeak over 65 minutes. The Gaussian mean $\mu$ is extracted and used to construct a calibration curve. (Right) Barium-133 spectrum from left panel calibrated with third degree polynomial fit, with calibration and test data collected from amp1.}
	\label{fig:fitSpectra_ba}
\end{figure}

\begin{figure}
	\centering
	\includegraphics[width=0.45\linewidth]{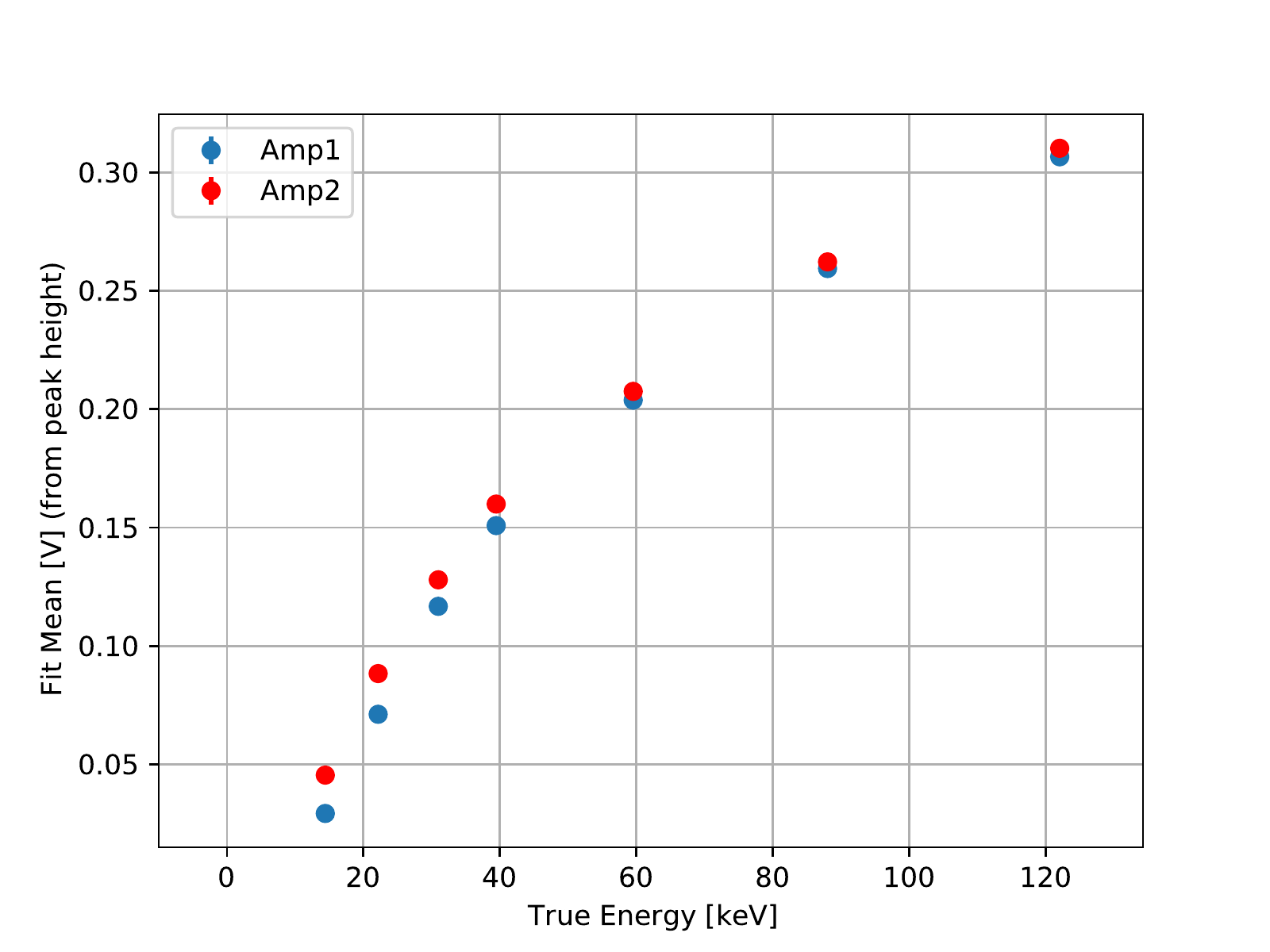}
	\includegraphics[width=0.45\linewidth]{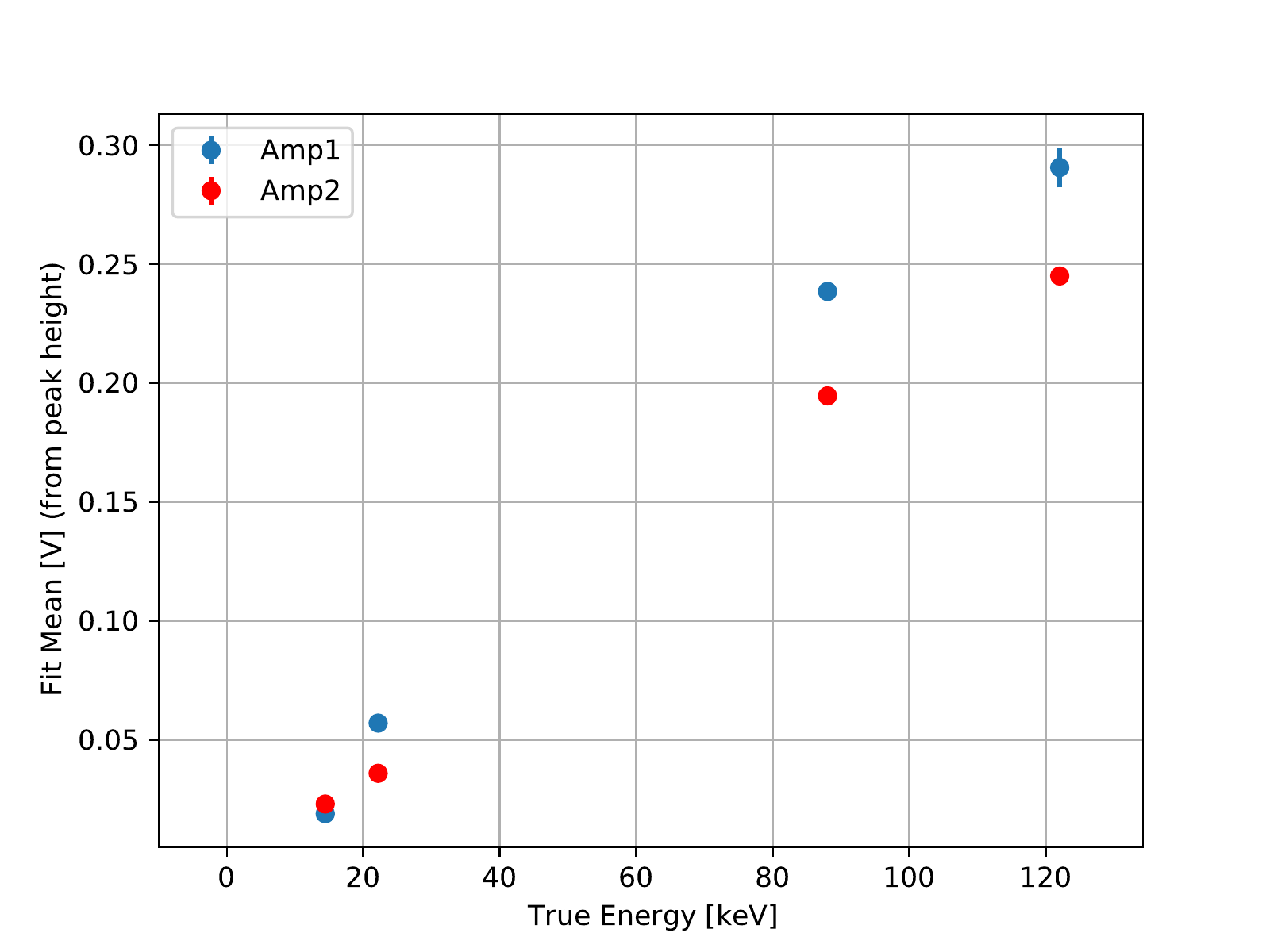}
	\caption{\apixone{} calibration curves for chip003 (left) and chip004 (right). Fewer data points were taken for chip004 due to time constraints, but only four points are required for calibration. Error bars from the error on the fit parameter fall within the diameter of the markers.}
	\label{fig:calib_noFit}
\end{figure}

The anticipated photopeaks (and single Compton edge) from the raw spectra collected from all the sources in Table~\ref{tab:sources} are fit with a Gaussian (see Fig.~\ref{fig:fitSpectra_ba} (left) for fitting to Barium-133) where the mean and width are extracted. Due to high background rates, photopeaks are identified and fit using only the five most populated bins. The Compton edge for the 122 keV emission line of Cobalt-57 at 39.46 keV is also considered, and fit with a modified Heaviside function (derivation in \citenum{inverseGaussian}). The Gaussian mean is used to create a calibration curve (see Fig.~\ref{fig:calib_noFit}). The response of \apixone{} to increasing charge is monotonic but not linear and the distribution is best fit with a third-degree polynomial function. 

A third degree polynomial describing the relationship between mV and keV the spectra to be calibrated (see Fig.~\ref{fig:fitSpectra_ba} (right)) with all photopeaks falling within 7\% of the anticipated value. Though this value does not yet meet \apix{} requirements, it is an impressive start for the first \apix{} design version. The measurement is limited by the analog readout method of data collection from a single $175\times175$~\um$^2$ pixel. Further improvement can also be expected as data analysis tools continue to develop.

\begin{figure}
	\centering
	\includegraphics[width=0.45\linewidth]{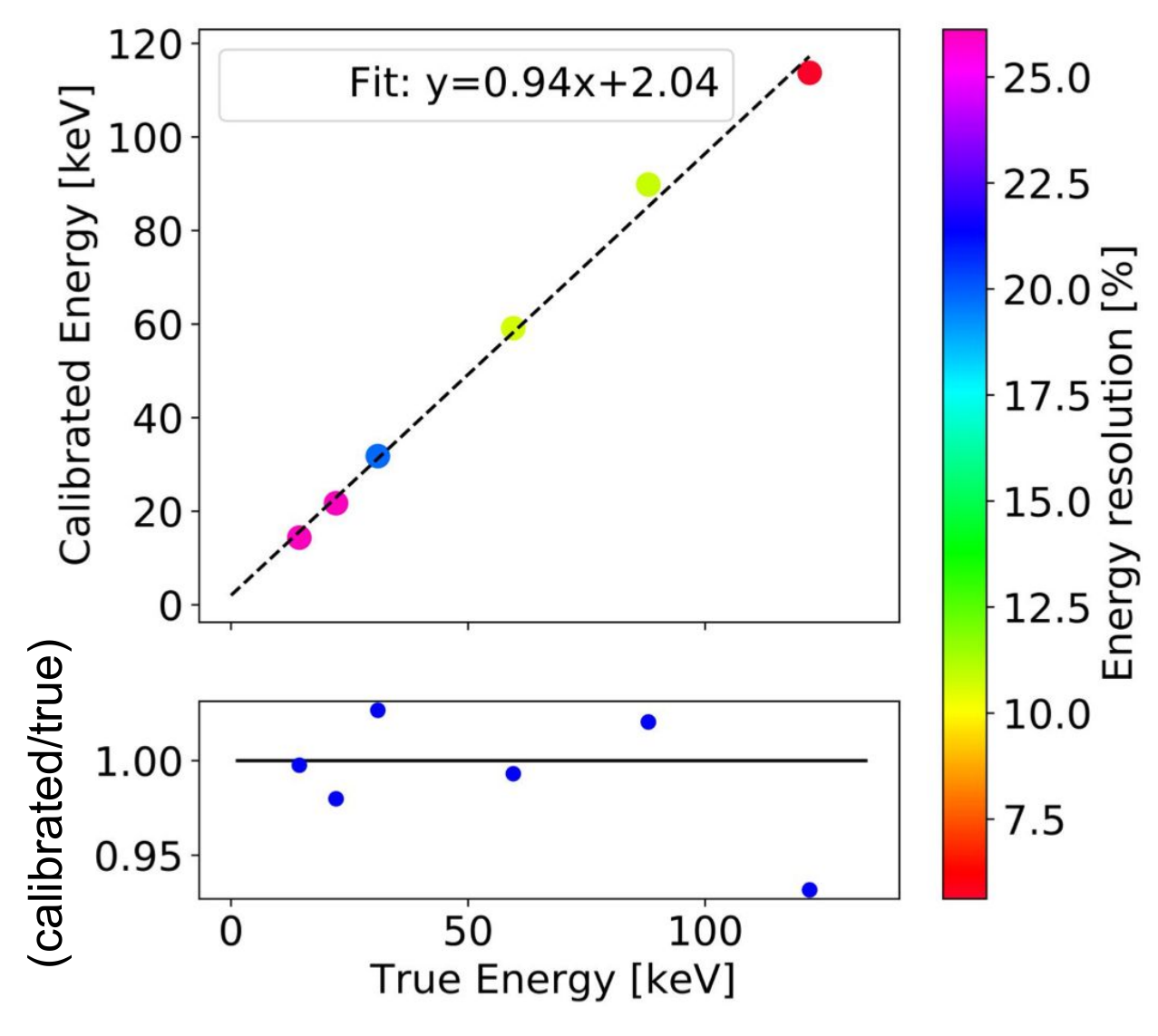}	
	\includegraphics[width=0.43\linewidth]{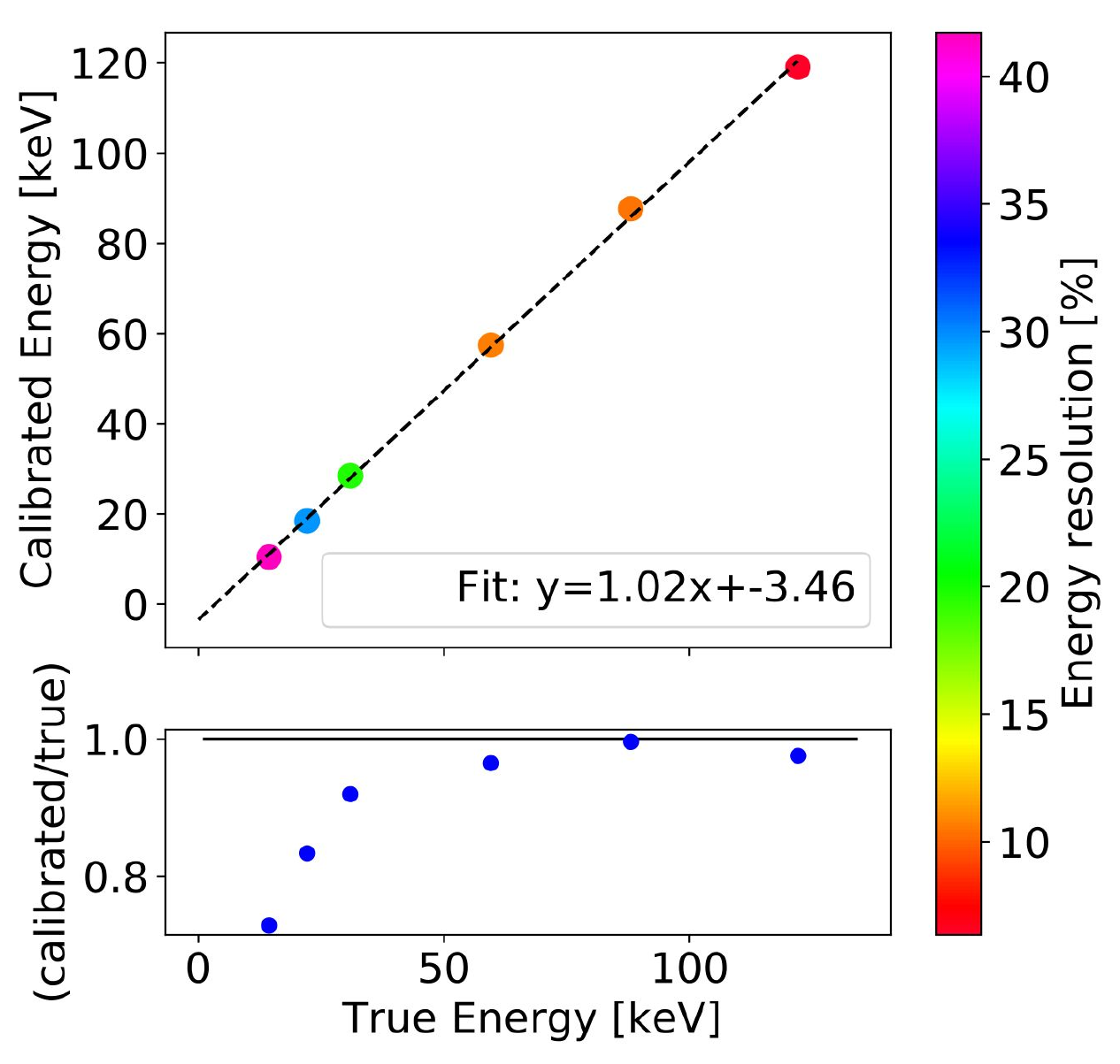}	
	\caption{Calibration performance of \apixone{} for all measured photopeaks. Calibration conducted with a third degree polynomial fit using test data collected by chip003 amp1 and calibrated by amp1 (left) and amp2 (right). Calibrating with the same data used to collect test data results in a calibrated mean energy that is closer to the expected value across the entire spectrum, and generally lower energy resolution.}
	\label{fig:calibratedEnRes}
\end{figure}

As the photopeak energy increases the energy resolution after calibration decreases (see Fig.~\ref{fig:calibratedEnRes}), indicating higher resolution. Since individual pixel responses differ (as shown in Fig.~\ref{fig:compareInjChips}), an individual calibration curve was constructed from data measured from each pixel. Figure~\ref{fig:calibratedEnRes} (right) illustrates the degredation in performance when a calibration curve derived from a different pixel is used. Figure~\ref{fig:fitSpectra_ba} (right) also illustrates the low energy threshold that \apixone{} can achieve. With the required noise cuts from Table~\ref{tab:noiseThreshold}, \apixone{} measures signal down to 13 keV.

\section{ASTROPIX$\_$V2}
\label{sec:v2}

\apixtwo{} follows \apixone{} part of the planned incremental design strategy. A key milestone achieved with \apixtwo{} is the ability to read and record digital data in addition to the analog data investigated with \apixone{} after the implementation of shielding against parasitic capacitance causing comparator oscillations. The results presented in this section consider only analog data in order to draw direct comparisons between ATLASPix and \apixone. Preliminary digital data conclusions will be presented in Section~\ref{sec:future}.

\subsection{Design Optimization and Setup for \apixtwo}
\label{sec:design_v2}

Pixel size is one of the incremental changes realized in \apixtwo, where the $175\times175$ \um$^2$ pixels of \apixone{} are enlarged to $250\times250$ \um$^2$. An \apixtwo{} chip is $1\times1$ cm$^2$ and contains a $35\times35$ array of these $250\times250$~\um$^2$ pixels. 

\apixtwo{} also features an updated guard ring design around these larger pixels to help control leakage and allow for a higher depletion voltage enabling a deeper depletion depth. By removing unused digital to analog converters (DACs) from the bias block, \apixtwo{} also sees a reduction of analog power from 14.7 mW/cm$^2$ to 3.4 mW/cm$^2$. Previously identified issues with crosstalk from \apixone{} were resolved in \apixtwo, and a global timestamp was enabled such that digital signal can be held in buffer during readout and read out with the subsequent trigger.

The test setup for \apixtwo{} is identical to that of \apixone, where the chip interfaces with a Nexys FPGA and GECCO board. \apixtwo{} is also mounted onto a carrier board, so the \apixone{} setup can be reused. Software updates have enabled Python wrappers to facilitate chip configuration and data-taking. Analog data can be collected from 35 pixels (one full row) of \apixtwo{}. This section focuses on the analog response of a single pixel located in the corner of the array. 

Like with \apixone{}, a bias voltage of -60V is supplied for these measurements. Preliminary studies indicate that maximal depletion of the sensor may occur around -160V, though this may still result in lower depletion than the design requirement of 500~\um. The true depletion depth and optimal bias voltage is still being studied. 

\subsection{\apixtwo{} Characterization Studies}

\subsubsection{Charge Injection Studies}
\label{sec:inj_v2}

As with \apixone, the response of \apixtwo{} was checked first using an injected charge. Updated DAC settings allow for the shaping of a shorter duration pulse (80~$\mu$s vs 600~$\mu$s) with a taller peak (0.41V vs 0.24V) than that of Fig.~\ref{fig:compareInjChips}. This shape allows for shorter deadtime and less pileup as compared to \apixone, but also uncovered a limitation in the analog \apixtwo{} design - an on-chip PMOS source follower saturates at 500 mV, serving as an upper limit to the height of analog pulse that can be collected. This causes the analog response of AstroPix v2 to saturate at energies of 90 keV - a factor of seven less than the required dynamic range of 25-700~keV. This will be resolved in future \apix{} versions, but must be considered in the analysis of \apixtwo. This limitation is only present in analog data and does not impact the performance of the digital output.


\subsubsection{Energy Calibration}
\label{sec:calib_v2}

\begin{figure}
	\centering
	\includegraphics[width=0.45\linewidth]{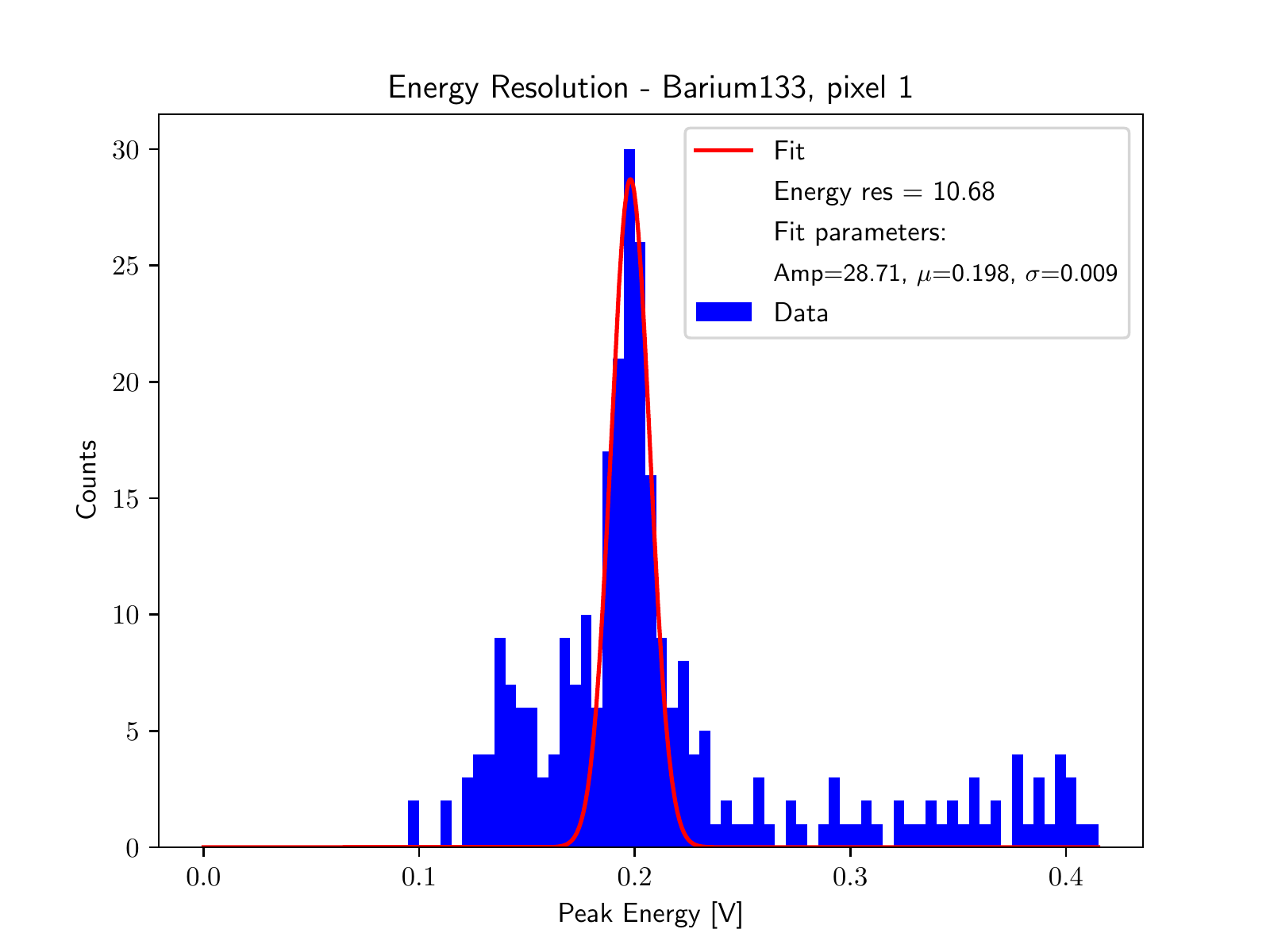}
	\includegraphics[width=0.45\linewidth]{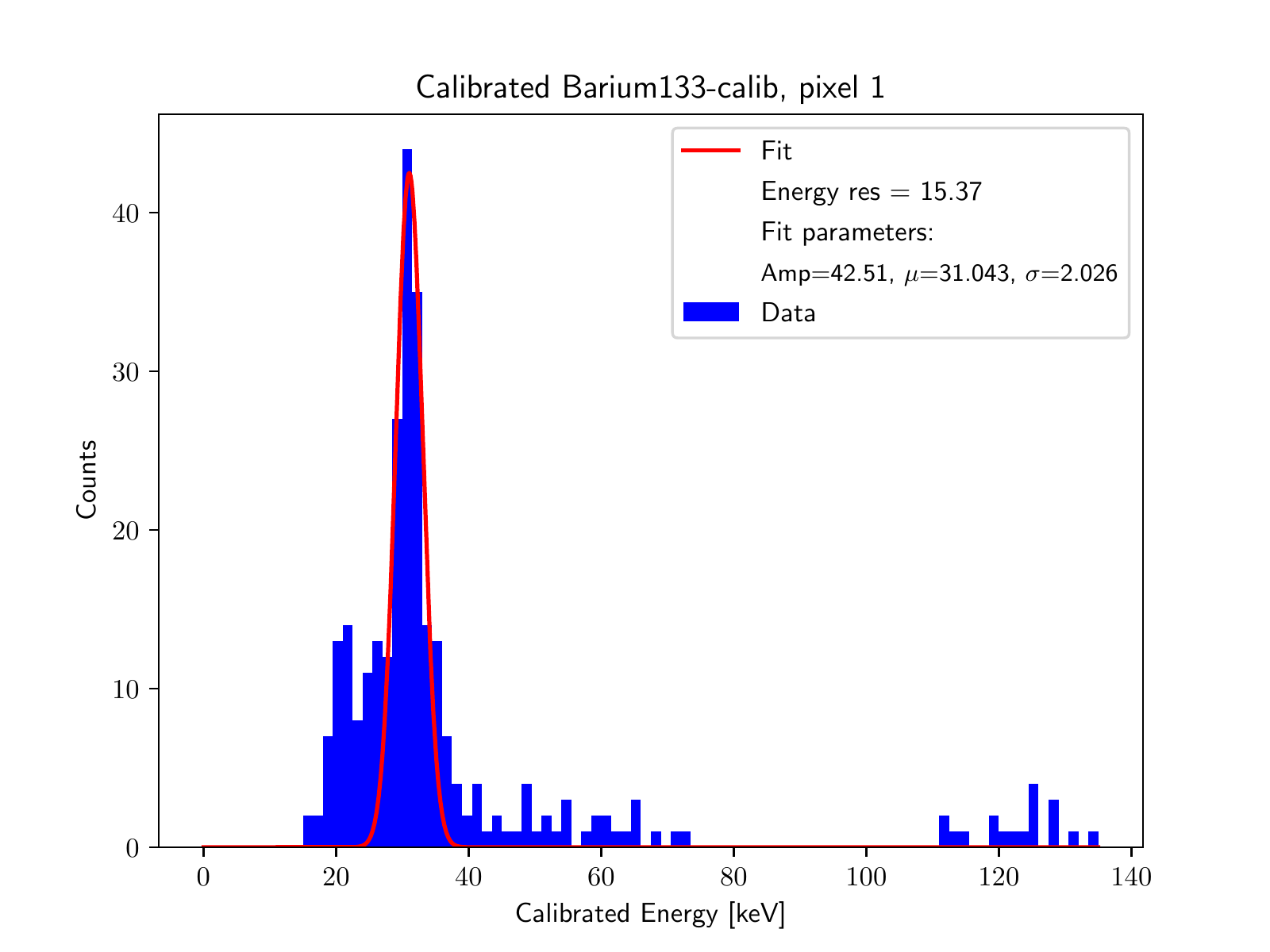}
	\caption{(Left) Gaussian fit to raw data collected with \apixtwo{} chip1 amp1 of Barium-133 over 5.5 hours. The Gaussian mean $\mu$ is extracted and used to construct a calibration curve. (Right) Barium-133 spectrum from left panel calibrated with cubic spline fit, with calibration and test data collected from amp1. The shape of the spectrum changes due to the nonlinearity of the calibration function.}
	\label{fig:fitSpectra_ba2}
\end{figure}

A calibration curve for \apixtwo{} was made using the response to radioactive isotopes with known emission lines (see Table~\ref{tab:sources}). An identical analysis strategy is used, where spectra of \apixtwo{} responses (the pulse height) are fit by a Gaussian whose parameters are correlated with anticipated photopeak energies. A higher threshold is required (see Table~\ref{tab:noiseThreshold}) due to increased electronics noise from the heightened capacitance related to the larger pixels. A spectrum is shown in Fig.~\ref{fig:fitSpectra_ba2} (left). Again, note the corresponding increase in measured mean as compared to \apixone{} (see Fig.~\ref{fig:fitSpectra_ba} (left)), yet smaller Gaussian standard deviation. The high rates of backscattered signal below the 30.97 keV photopeak noted in \apixone{} is also decreased in \apixtwo{} possibly due to a larger distance between the source and the sensor (reducing background from low-energy scattered \gam s) and/or more complete charge collection due to a larger depletion region.

\begin{figure}
\centering
\begin{minipage}{0.47\textwidth}
  \centering
	\includegraphics[width=\linewidth]{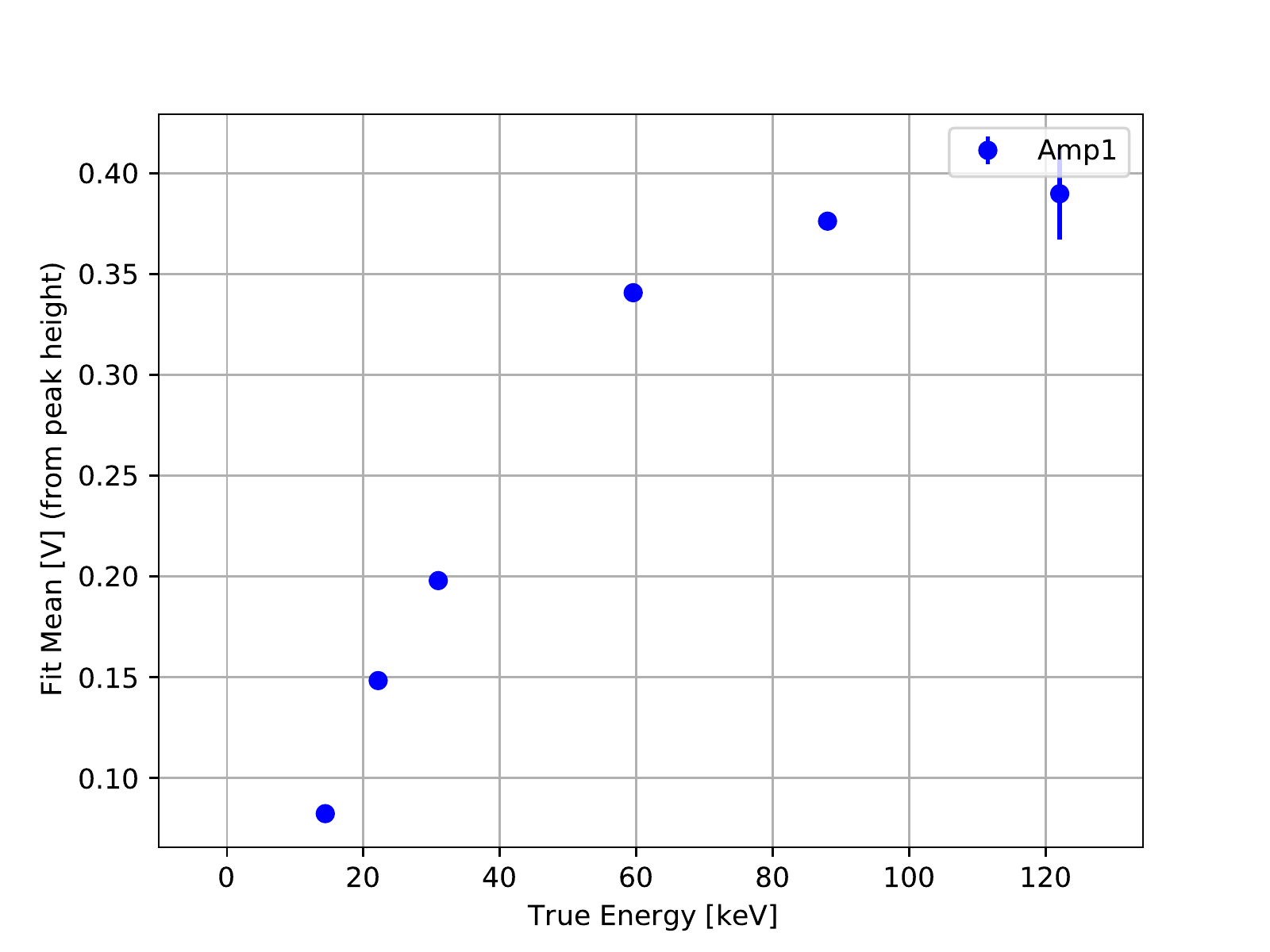}
	\captionof{figure}{\apixtwo{} analog calibration curve for chip1. Most error bars fall within the diameter of the markers. The impact of PMOS source follower saturation beginning around 350 mV is evident.}
	\label{fig:calib_noFit2}
\end{minipage}%
\begin{minipage}{0.06\textwidth}
	~
\end{minipage}%
\begin{minipage}{0.47\textwidth}
  \centering
	\includegraphics[width=\linewidth]{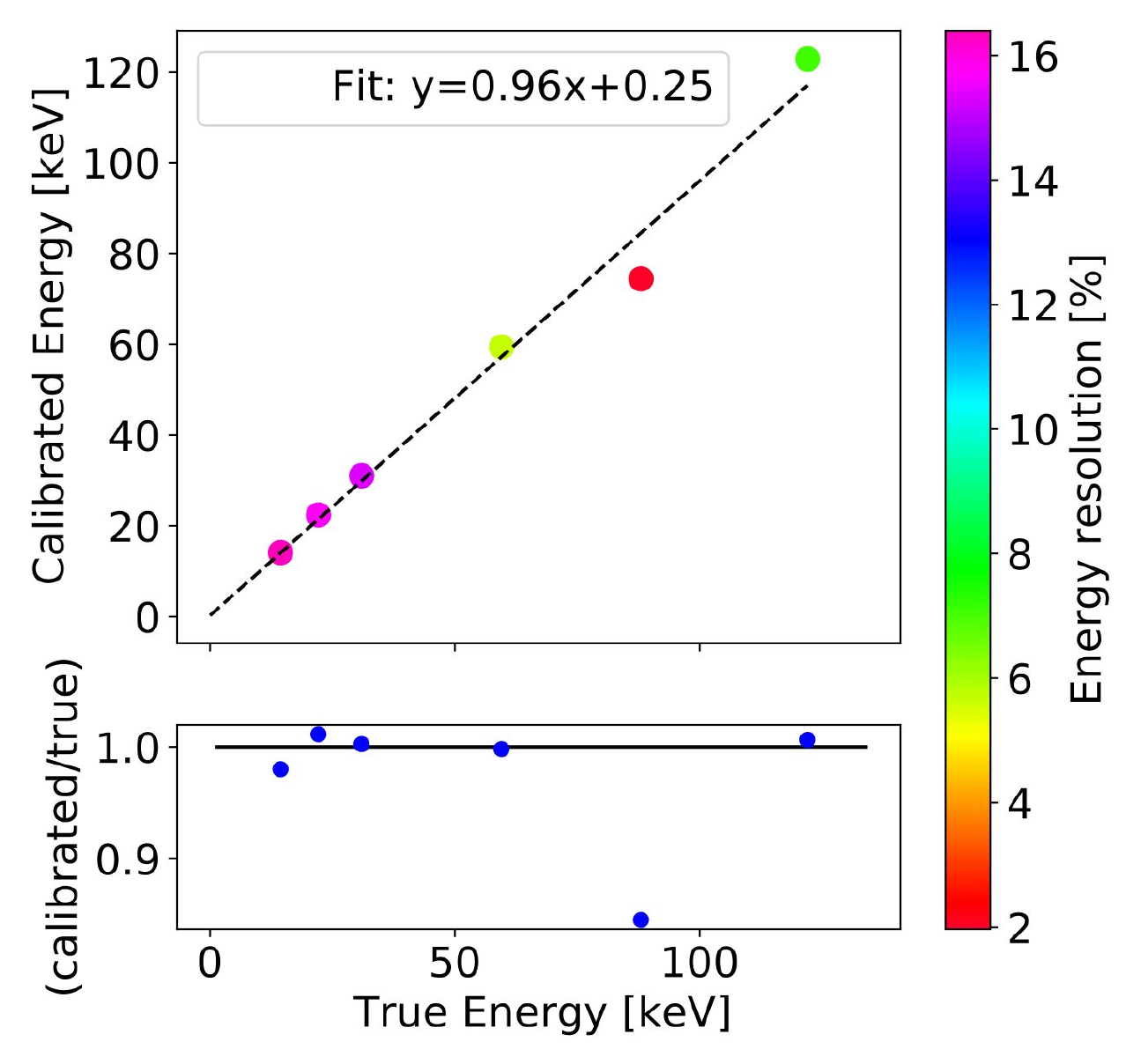}	
	\captionof{figure}{Calibration performance for all tested photopeaks. Calibration conducted with a cubic spline fit using test data collected by chip1 amp1.}
	\label{fig:calibratedEnRes2}
\end{minipage}
\end{figure}

Like \apixone{}, the analog response of \apixtwo{} to increasing charge is monotonic but not linear. The impact of the PMOS source follower saturation becomes evident around 60 keV and inhibitive around 90 keV as the response pulse height rises more slowly and asymptoticly approaches a value around 400 mV. This impact is large in analog data analysis, however the analog studies presented in this work are intended to serve as a preliminary assessment of the chip's capabilities. Digital readout will be utilized by future \apix{} versions (with a first look presented in Section~\ref{sec:future}), and it is not impacted by this saturation.

Due to the source follower saturation, the analog calibration curve (see Fig.~\ref{fig:calib_noFit2}) is best fit with a cubic spline fit. Resulting calibrated spectra (see Fig.~\ref{fig:fitSpectra_ba2} (right)) reproduce photopeaks falling within 15\% of the anticipated value. This includes the outlying 88 keV photopeak from Cadmium-109. A lack of statistics for the raw spectrum led to poor fitting which propagated to the nonlinearly scaled calibrated spectrum. Without consideration of this point, the cubic spline calibrations agree with expected values within 3\%. This is a vast improvement over \apixone{} calibration, and provides a point of comparison for eventual studies of the digital output. A higher degree of confidence could be placed on a calibration function spanning only from 14 - 60 keV, in the range that \apixtwo{} is unaffected by source follower saturation. Figure~\ref{fig:fitSpectra_ba2} (right) also illustrates that the low energy threshold of 13 keV is consistent with that of \apixone.

As the photopeak energy increases the energy resolution after calibration also increases (see Fig.~\ref{fig:calibratedEnRes2}). Compared to \apixone{} (see Fig.~\ref{fig:calibratedEnRes}), \apixtwo{} measures a better energy resolution for each calibrated point with $E_{res}<16\%$ for all points as compared to $E_{res}<25\%$ with \apixone.

The case study of the 30.97 keV photopeak of Barium-133 is considered in Table~\ref{tab:compareRequirements}. Analog measurements from \apixtwo{} perform better than those from \apixone, and the planned incremental upgrades to \apix{} design is delivering results consistently closer to design requirements and with higher precision. It is assumed that the final \apix{} design will be dominated by electronics noise from the range of 25-122 keV with a constant energy resolution of 5 keV RMS. Therefore the required energy resolution performance at 30.97 keV of 38\% is already being met with analog measurements made in the lab of all tested \apix versions. It must be noted, though, that the presented measurements are the result of analog measurements of a single, well-characterized pixel and ultimate performance of \apix{} must be judged on the digital performance of the full array.

The lowest measured energy resolution with ATLASPix is also included in Table~\ref{tab:compareRequirements}, although the thinner ATLASPix wafer is easier to fully deplete and was previously optimized by the ATLASPix team for measurement while \apix{} characterization and optimization is occurring in tandem.
\begin{table}[b!]
	\caption{Comparison of \apix{} measurements with ATLAPix performance and AstroPix requirements. \apix{} energy resolution requirement is 5 keV RMS at 122 keV which is assumed to be constant down to 25 keV, or a 38\% energy resolution at 30.97 keV. Energy resolution values are stated with respect to the FWHM. The 500~\um{} wafer requirement represents the depletion depth requirement.}
	\label{tab:compareRequirements} 
	\centering
	\begin{tabular}{|l|c|c|}\hline
		&Energy Resolution at 30.97 keV [\%]& Wafer Thickness [\um]\\\hline
		Requirement& $38\%$&500 \\
		ATLASPix& $7.3 \pm 1.2$&100\\
		\apixone{}& $19.9 \pm 7.4$&700\\
		\apixtwo{}& $15.4 \pm 2.9$&700\\
		\hline
	\end{tabular}
\end{table} 

When considering calibration as a whole, \apixtwo{} experiences limitations in its calibration power due largely to the PMOS source follower saturation at 500 mV. The effective range of calibration from 0-60 keV is more limited than that of \apixone{} which could successfully calibrate up to 122 keV. However, measured points within this valid range were better calibrated with \apixtwo{} as compared to \apixone. This effect is noted and will improve with updated hardware in future \apix{} iterations, however a shift from analog data to digital data will also limit the scope of this setback. Digital data, not analog data, will be utilized exclusively in the final \apix{} design so shortcomings in the analog circuitry (used at this stage for testing) do not necessarily inhibit the overall intended chip performance.

\section{Radiation Testing}
\label{sec:rad}

Preliminary testing of \apixtwo{} has been conducted in two separate beam environments. The intention of this testing is to raise the technology readiness level (TRL) of \apix{} and increase confidence in the chip's capabilities on orbit. Results presented here are preliminary, and currently undergoing independent assessment.

\apixtwo{} participated in two campaigns at the Fermilab Test Beam Facility (Ref.~\citenum{ftbf}) where it was exposed to a 300 kHz 120~GeV proton beam. Both the digital and analog performance was assessed in these qualitative tests. \apixtwo{} was able to register and record signals in this high-energy and high-flux environment, though the rate of interaction is much greater than the environment \apix{} is designed for in orbit. This high rate overwhelmed the current software designed for debugging and chip characterization, which optimized for lower rates of $<$5 Hz and is not the final flight-ready version. None of the monitored power rails increased current draw during beam running as compared to bench running. These successful tests were the first radiation tests of any \apix{} chip and serve as an important milestone for the continued development of the technology. 

\apixtwo{} was also subjected to radiation testing at the 88-Inch Cyclotron operated in conjunction with the Berkeley Accelerator Space Effects facility at Lawrence Berkeley National Laboratory (Ref.~\citenum{lbnl}). Here, heavy ions ranging from Argon to Xenon with an atomic tune of 16 MeV/(atomic mass unit) illuminated \apixtwo. The susceptibility to latchup was first recorded. This is a state of inactivity in which an incident ion deposits significant charge in a pixel, inducing a perpetually open transistor and leading to runaway current draws. The latchup state must be corrected through a chip reset and reconfiguration. The sensor successfully survived this test, withstanding a fluence (integrated flux) of $1\times10^7$ particles/cm$^2$ at an effective linear energy transfer (LET) of 65 MeV*cm$^2$/mg. 

Testing for single event functional interrupts (SEFI) was also conducted. This event may occur when a heavy ion collision interacts with the digital periphery and flips a bit upon readout causing data degradation or corrupted configuration. Proper operation can be restored with a chip reset or reconfiguration as no permanent damage is sustained. Possible evidence of SEFIs with an onset of LET=22.5 MeV*cm$^2$/mg were noted during the run, identifying elements of the digital framework of the chip that may be susceptible. This informed an updated design of future \apix{} versions, which can limit the onset of SEFIs to higher LET values.

\section{Future Plans and Summary}
\label{sec:future}

The digital output of \apixtwo{} is still being tested, and will be included in a dedicated paper. The previously considered analog data is the signal pulse returned from a single pixel. For digital output, this pulse is digitized on chip and a bitstream with encoded hit information is read out. A particle interaction in a pixel will cause the pixel's comparator to trigger. The comparator also calculates a Time over Threshold (ToT) value associated with the particle interaction. This is a true time in microseconds that the response pulse exceeded a defined threshold value. This ToT value, rather than pulse height as seen in the analog case, serves as a proxy for the amount of charge deposited. A larger ToT value indicates a larger charge deposition. The threshold value used to calculate the ToT is universal for every pixel across the chip and in \apixtwo{} cannot be set on a pixel-by-pixel basis. All comparator readings from the full array are OR'd into one channel containing row hit information and another of column hit information. The triggering of any comparator will impact an interrupt signal which indicates that the full chip should be read out. 

\begin{figure}
	\centering
	\includegraphics[width=0.5\linewidth]{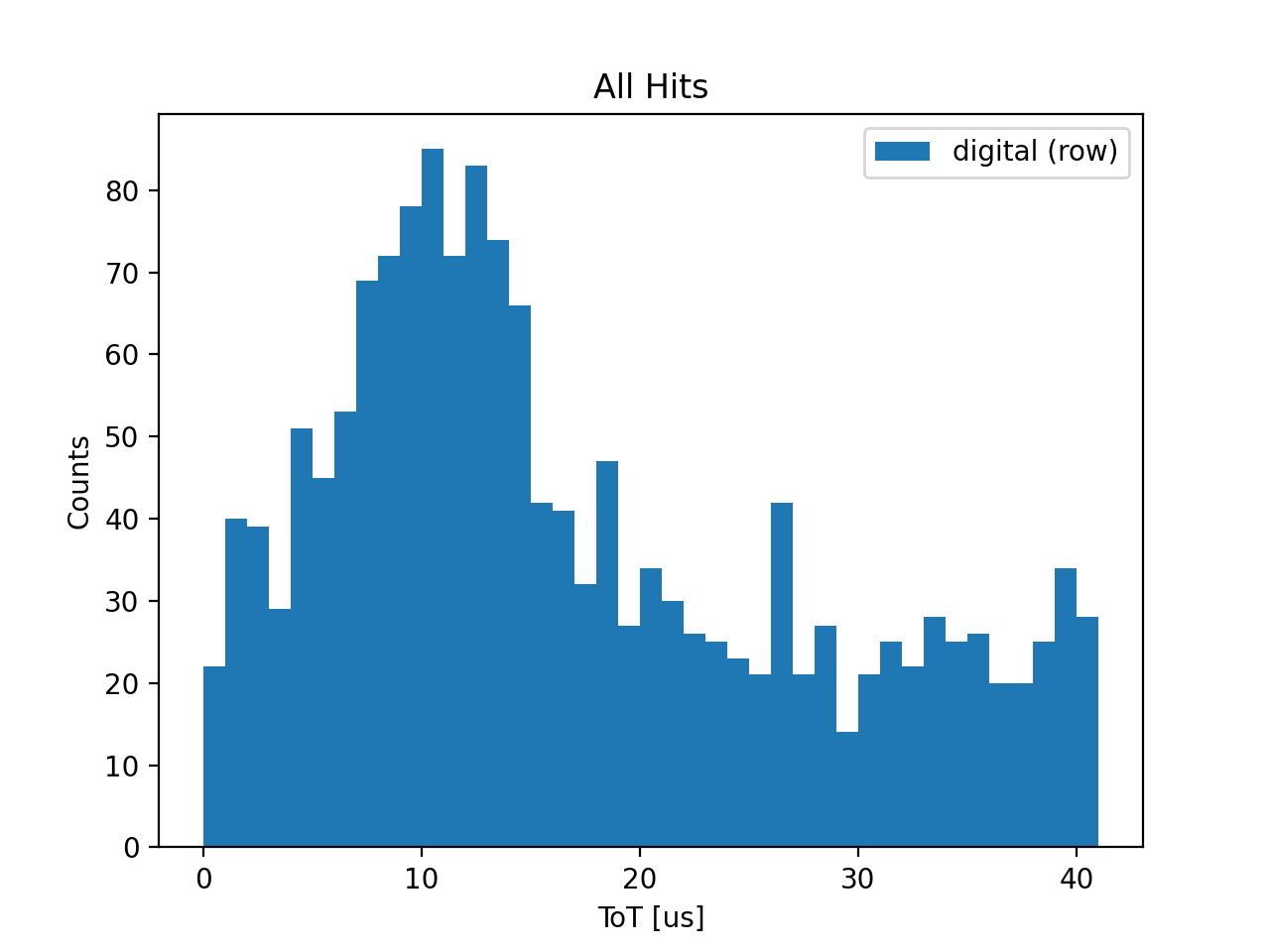}	
	\caption{Preliminary spectrum of digital data recorded over 3 hours from one \apixtwo{} pixel with a Barium-133 source.}
	\label{fig:digitalSpectrum}
\end{figure}

Preliminary work shows that the sensor is able to trigger upon, read out, and decode data (see Fig.~\ref{fig:digitalSpectrum}). A clear photopeak in response to a Barium-133 source can be seen in this ToT spectrum when data from a single digital pixel is read out. This early work is intended to serve as an illustration that digital data can be collected and interpreted, and not yet serve as an indication of \apixtwo{} digital performance. Work is currently underway to further understand the performance of the digital readout system with studies planned to map noise and gain variations between pixels, record digital signals from more pixels on the array, and optimize the comparator's threshold value.

\apixthree{} was submitted for fabrication in Summer 2022. Additional power-reducing improvements will be incorporated into the design with a target total draw of $<1$mW per cm$^2$. The pixel size will also continue to increasing in \apixthree, up to $500\times500$~\um$^2$. Within this pixel pitch, the active area will be $300\times300$~\um$^2$ and the total array will increase to $2\times2$~cm$^2$. This additional space allows for the removal of conductive $p$-wells between pixels, limiting noted leakage of the charge cloud into the array periphery. \apixthree{} will also be designed with the possibility to dice the wafer into a quad-chip - four individual chips cut from the wafer as one $4\times4$ cm$^2$ square of identical MAPS chips, quadrupling the number of pixels. This is the unit that will act as a building block for larger structures such as a future telescope tracker. Readout capability of the quad chip involving daisy-chaining the signal from one chip through the others on SPI readout was implemented in \apixtwo, but will be fully tested in \apixthree. With this full quad-chip, environmental testing of of \apixthree{} in a space environment is also planned with the development of a four-layer hosted payload to be included on a sounding rocket mission. The success of this testing will further raise the TRL. In AstroPix$\_$v4, digital state triplication will be introduced to protect the chip against single event functional interrupts possibly observed in the radiation testing discussed in Sec.~\ref{sec:rad}. 

In summary, \apix{} is a CMOS MAPS chip designed for the space environment. A robust design and characterization strategy is in place to advance the technology from its current stage to flight readiness. The current version under test, \apixtwo, is capable of analog and digital readout and a comprehensive analysis of the analog response has been presented and compared to the performance of the previous version, \apixone. Metrics such as total power draw do not yet meet \apix{} design requirements, but the \apix{} design strategy relies upon incremental changes between versions building to the final design. Therefore, the consistent improvement of performance in updated versions is a success toward realizing a final \apix. The utilization of \apix{} in future \gam{} telescopes (such as \amx, as described in Ref.~\citenum{amx}) will enable high energy, angular, and position resolution for low power consumption these next-generation detectors. 


\acknowledgments 
 
The authors would like to thank the input of the \apix{} team for their help and support in development, fabrication, firmware/software production, and testing of \apix. 

This work is funded in part by 18-APRA18-0084. HF and MN acknowledge support by NASA under award number 80GSFC21M0002. AS was supported by the National Aeronautics and Space Administration (NASA) through a contract with ORAU. The views and conclusions contained in this document are those of the authors and should not be interpreted as representing the official policies, either expressed or implied, of the National Aeronautics and Space Administration (NASA) or the U.S. Government. The U.S. Government is authorized to reproduce and distribute reprints for Government purposes notwithstanding any copyright notation herein.

\bibliography{astropix} 
\bibliographystyle{spiebib} 

\end{document}